\documentclass[twocolumn,twoside,preprintnumbers,amsmath,amssymb,showkeys]{revtex4}

\usepackage{epsfig}

\usepackage{graphicx}

\usepackage{fancyhdr}
\usepackage{pslatex}

\begin{document}
\title {What to Expect When You're Expecting: Femtoscopy at the LHC}

\author{Mike Lisa$^1$}

\affiliation{$^1$Department of Physics, Ohio State University, 1040 Physics Research Building, 191 West Woodruff Ave., Columbus, OH 43210, USA}

\begin{abstract}
A huge systematics of femtoscopic measurements have been used over the past 20 years to
characterize the system created in heavy ion collisions.  These measurements cover two orders of magnitude
in energy, and
with LHC beams imminent, this range will be extended by more than another order of magnitude.
Here, I discuss theoretical expectations of femtoscopy of $A+A$ and $p+p$ collisions at the LHC, based on
Boltzmann and hydrodynamic calculations, as well as on naive extrapolation of existing systematics.
\end{abstract}

\vskip -1.35cm
\maketitle
\thispagestyle{fancy}
\setcounter{page}{1}
\bigskip

\section{Introduction}

What distinguishes ultrarelativistic heavy ion physics from particle physics is its focus on geometrically large systems.
The desire is not to understand fundamental processes, as in the latter field, but to create and probe new states of matter
and access the only phase transition associated with a fundamental interaction (QCD).
The geometrically-sensitive, bulk properties are the crucial ones, and these are reflected in the soft ($p_T \sim \Lambda_{QCD}$)
sector.

In soft sector observables, long-term baselines have been established over a large energy range.  Prior to first data at RHIC,
it was commonly speculated (and hoped) that large deviations from these systematics (e.g. $\pi^0/\pi^\pm$ ratios, sidewards
flow, strangeness enhancement, total multiplicity) would signal clearly the qualitatively different nature of the system created
there~\cite{Harris:1996zx}.  In femtoscopic systems, rather generic arguments led
to expectations~\cite{Rischke:1996em,Bass:1999zq} of a rapid increase, with $\sqrt{s_{NN}}$,
in the pion ``HBT radii'' $R_{out}$ and $R_{long}$, reflecting relatively long timescales of the transition
from deconfined QGP to confined hadronic matter.

Such dramatic speculations are largely absent today, in anticipation of LHC collisions.
Soft-sector, global observables at RHIC are only quantitatively different than they are at lower energies.
Even in the high-$p_T$ sector, where
jet suppression and partonic energy loss measurements have generated huge excitement, energy scans
at RHIC reveal that the data indicate more of an evolution than a revolution.

This is all to the good.
Discoveries via sharp jumps {\it \`{a} la} superconductivity are not our lot.
The real science behind heavy ion measurements (at very high energies as well as at much lower ones)
lies in understanding the {\it details} of the data.

Femtoscopy~\cite{Lednicky:1990pu}, the geometric measurement of systems at the fermi scale, has been used
to characterize the freezeout substructure of heavy ion collisions for two decades in time, and over two
decades in collision energy~\cite{Lisa:2005dd}.  Soon, this energy range will be
extended by another decade at the LHC.

\begin{figure}[t!]
{\centerline{\includegraphics[width=0.45\textwidth]{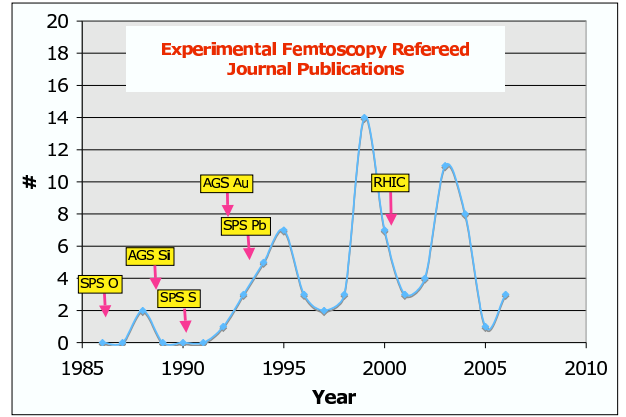}}}
\caption{
\label{fig:PapersPerYear}
The number of refereed-journal publications reporting new femtoscopy results in relativistic
heavy ion collisions.  Beginning of availability of new beams are indicated in yellow boxes.
}
\end{figure}

It will be important to understand the evolution of the 
non-trivial space-time substructure of the bulk system as the initial conditions change dramatically
with energy.
In Figure~\ref{fig:PapersPerYear} is shown the number of refereed-journal
papers of experimental femtoscopic results in relativistic heavy ion collisions, as a function of publication year.
Even armed with nothing more than this Figure and Reinhard Stock's observation~\cite{Stock:2004cf}
that ``HBT experiences a renaissance of new insights roughly every five years,'' we may confidently
expect a barrage of new femtoscopic information to digest within the next 5 years.

Here, we discuss predictions for femtoscopy at the LHC.  In the next Section, we consider the case
of simple extrapolation of measured femtoscopic trends, with no reference to physics per se.  In Section~\ref{sec:cascade},
we consider predictions of Boltzmann/cascade transport calculations, and in Section~\ref{sec:hydro} those
of hydrodynamical models.  Finally, we discuss speculations on the physics behind femtoscopic measurements in
$p+p$ collisions, which will, in fact, be the first results available at the LHC.  At the end we summarize.

\vspace*{-3mm}

\section{Nothing New Under the Sun (NNUS) Scenario}
\label{sec:NNUS}

\vspace*{-3mm}

Femtoscopic measurements display rich, multidimensional and nontrivial systematic dependences
upon kinematic ($p_T$, $y$, etc) variables and particle species~\cite{Lisa:2005dd,Lisa:2005js}.
The dependence upon {\it global} variables such as $\sqrt{s_{NN}}$ and impact parameter, however,
appears significantly more trivial.  Schematically characterizing the measured femtoscopic length
scales as a multidimensional function, evidence thus far indicates an overall factorization
\begin{eqnarray}
\label{eq:Factorization}
R\left(\sqrt{s_{NN}},A,B,|\vec{b}|,\phi,y,m_T,m_1,m_2\right)  \nonumber \\
 = R_{g}\left(\sqrt{s_{NN}},A,B,|\vec{b}|\right)\cdot F_{k}\left(\phi,y,m_T,m_1,m_2\right) \\
 = R_{g}\left(M\right)\cdot F_{k}\left(\phi,y,m_T,m_1,m_2\right) , \nonumber
\end{eqnarray}
where $F_k$ is a dimensionless function containing, e.g. decreasing ``HBT radii'' with particle $m_T$.
The dimensional scale $R_g$ is determined by global observables.  However, as indicated by
the second equality of Equation~\ref{eq:Factorization}, to good approximation the only relevant global
observable is the total multiplicity $M$ of the collision.
In fact, this multiplicity dominance well apply to all soft-sector observables~\cite{Caines:2006if}.

\begin{figure}[t]
{\centerline{\includegraphics[width=0.42\textwidth]{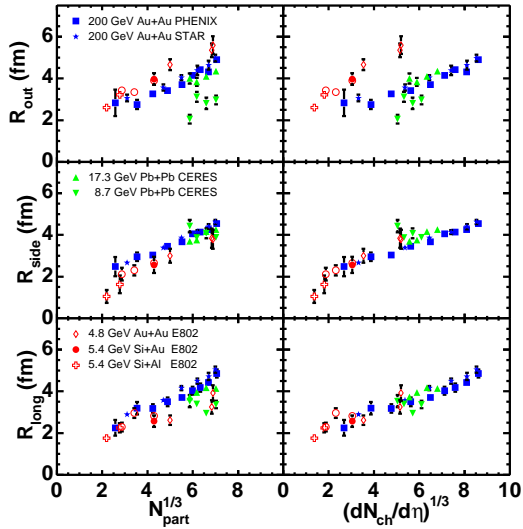}}}
\caption{Pion HBT radii plotted versus the number of participating nucleons (left)
and versus the charged particle multiplicity (right) for collisions of varying centrality
and a wide range of energies.  Compilation from~\cite{Lisa:2005dd}.
\label{fig:RversusMultAnnRev}}
\end{figure}

There are at least three caveats to the above statement.
Firstly,
the CERES~\cite{Adamova:2002ff} collaboration
has shown that the scale depends also on the freeze-out chemistry (baryon-to-meson ratio) in
addition to the multiplicity.  This is important at low (AGS) energies.
For collisions above top SPS energy, $\sqrt{s_{NN}}\sim 17$~GeV, the chemical evolution is
sufficiently weak that one may consider multiplicity only.
Secondly, as seen in Figure~\ref{fig:RversusMultAnnRev}, there is some residual dependence of the outward radius
on $\sqrt{s_{NN}}$ in addition to multiplicity; in this Section, we ignore this potentially
important detail.
Finally, the azimuthal ($\phi_p$, as determined relative to the reaction plane)
dependence~\cite{Lisa:2000ip,Lisa:2000xj,Wells:2002phd,Adams:2003ra}
likely at some point violates the factorization.  While it has not been experimentally tested,
two collisions producing the same
multiplicity, one very peripheral (i.e. spatially anisotropic in the entrance channel)
at high energy and the other
very central at low energy, presumably generate freezeout distributions with different
spatial anisotropy, which is then reflected in the azimuthally-sensitive femtoscopy~\cite{Retiere:2003kf}.
See Section~\ref{sec:hydro} for further discussion.

These caveats stated, however, the factorization of Equation~\ref{eq:Factorization} is probably
our best, zero-new-physics guide to simple extrapolation of femtoscopic trends measured over two
orders of magnitude in $\sqrt{s_{NN}}$ and from from the lightest ($p+p$) to the heaviest ($Pb+Pb$)
systems.  
Figure~\ref{fig:RversusMultAnnRev} suggests a simple form $R_g(M) \propto M^{1/3}$; this ignores the finite
offset $\sim 1$~fm when extrapolating $M \rightarrow 0$, but this is negligible for high multiplicity.
This relation may reflect that a constant freezeout density drives the femtoscopic scales~\cite{Adamova:2002ff},
though this neglects any dynamic effects.
Assuming that this simple proportionality continues, then,
we know $R_g(M)$ and 
determining femtoscopic expectations boils down to anticipating the multiplicity at the LHC.

A naive extrapolation~\cite{Lisa:2005js,Caines:2006if} of systematics suggests that
$dN/dy$ at the LHC will be 60\% larger than that observed at RHIC.  Thus, the zeroth-order
expectation is that length scales at the LHC will be 17\% ($1.6^{1/3}=1.17$) larger than
those measured at RHIC, for all kinematic selections and particle species, according
to Equation~\ref{eq:Factorization}.

Going beyond simple extrapolation to include a physical picture, saturation-based
calculations~\cite{Kharzeev:2004if} give much higher multiplicity-- roughly triple
that at RHIC.  This leads to expectations of length scales 45\% higher than those
at RHIC.  Thus, $R_{long}$ for pions at midrapidity and low $p_T$ in central collisions
would be $1.45\times 7~{\rm fm} = 10~{\rm fm}$.



Multiplicity predictions based on Boltzmann/cascade calculations can be significantly higher yet.
Selecting two for which femtoscopic predictions also exist (Section~\ref{sec:cascade}),
A Multi-Phase Transport (AMPT) calculation~\cite{KoWPCF06}
and the Hadronic Rescattering Model (HRM)~\cite{Humanic:2005ye} predict $5\times$ and $7\times$ RHIC
multiplicity, respectively.  Thus, femtoscopic scales at LHC may be as much as 90\% higher
than at RHIC.  Depending on the final-state interaction which produces the two-particle
correlation function, measuring length scales of $\sim 15$~fm may challenge experimental
two-track resolutions.  For two-pion correlations, such scales are within the capabilities
of the ALICE detector~\cite{Alessandro:2006yt}.

\vspace*{-3mm}

\section{Boltzmann Transport Calculations}
\label{sec:cascade}

\vspace*{-3mm}

More interesting than simple scaling relations are models with real physics and
dynamics, such as transport calculations.
Boltzmann/cascade transport models generally reproduce ``HBT radii'' at RHIC
better than do hydrodynamic calculations~\cite{Lisa:2005dd}.  
The reasons behind this include different 
physics in the models, a more detailed description of the kinetic freezeout,
and the use of more appropriate methods of calculating the radii~\cite{Frodermann:2006sp}.
Predictions of pion HBT radii with each 
of the transport calculations discussed in Section~\ref{sec:NNUS} reveal predictions
more subtle than the simple multiplicity-scaling discussed above.

\begin{figure}[t]
{\centerline{\includegraphics[width=0.45\textwidth]{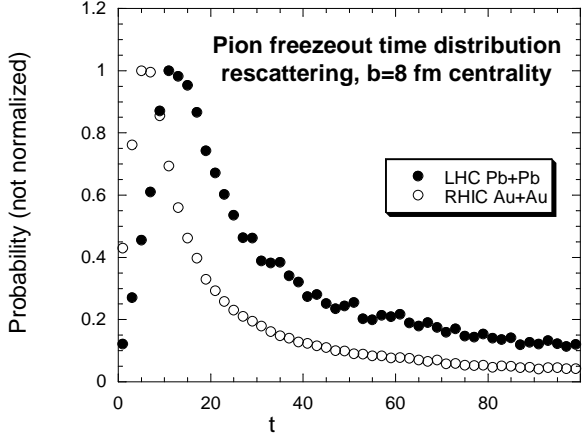}}}
\caption{
\label{fig:HRMtime}
The freezeout time distribution from the Hadronic Rescattering Model of Humanic~\cite{Humanic:2005ye} for RHIC and LHC conditions.
}
\end{figure}

\begin{figure}[t!]
{\centerline{\includegraphics[width=0.45\textwidth]{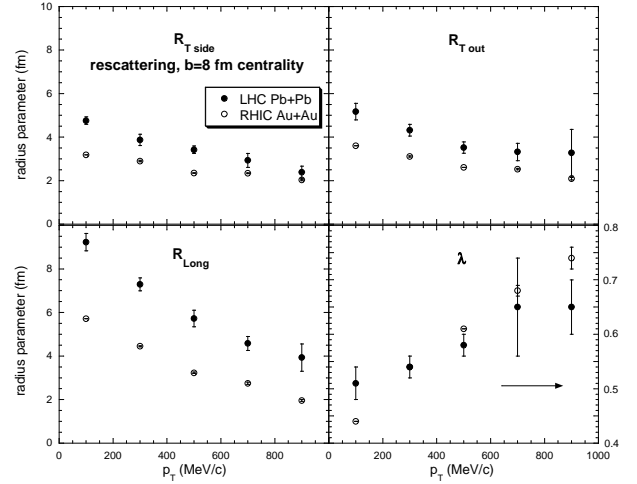}}}
\caption{
\label{fig:HRMradii}
HBT radii from fits to pion correlation functions from the
Hadronic Rescattering Model of Humanic~\cite{Humanic:2005ye} for RHIC and LHC conditions.
}
\end{figure}

For an infinite and boost-invariant system (only an approximation of reality, of course),
the longitudinal HBT radius $R_{long}$ is proportional to the system evolution
time (i.e. between interpenetration of the ions and kinematic freezeout of the
products)~\cite{Akkelin:1995gh,Lisa:2005dd}.  Naturally, this is not a unique, system-wide
time, but a distribution.  An example is seen in Figure~\ref{fig:HRMtime}, in which
the pion freeze-out time distribution for collisions at RHIC and LHC are compared in
the HRM calculation.  The LHC timescales are roughly double those at RHIC.
Although HRM is not explicitly a boost-invariant model, we see in Figure~\ref{fig:HRMradii}
that $R_{long}$ reflects this timescale increase, roughly doubling when the energy
is increased from RHIC to LHC energies.
The $\sim 70\%$ increase in $R_{long}$ is roughly consistent, then, with expectations from
both a timescale and from the multiplicity-scaling point of view.
This is certainly not a coincidence, as the increased timescale is due in large part to the increased multiplicity.

On the other hand, there is more going on.  The predicted $p_T$-dependence of 
both $R_{long}$ and $R_{side}$ are steeper at the LHC than at RHIC.  Also, the
increase in $R_{side}$ is significantly less than 90\%.  Both of these effects
are consistent with a freezeout scenario with significantly increased transverse
flow~\cite{Retiere:2003kf}.  Indeed, transverse momentum distributions predicted by HRM
are significantly harder (less steep) at the LHC than those at RHIC.  
Since the $p_T$ dependence of HBT radii~\cite{Lisa:2005dd} and spectra  
in the soft sector are observed to
change very little between $\sqrt{s_{NN}} = 20 \div 200$~GeV, it will
be interesting to see whether this trend is broken at the LHC, as predicted by HRM.

The HRM model is a deliberate effort to use the simplest (often criticized as too simplistic)
physics picture, free of novel phases like QGP.  It is a pure hadron-based transport calculation,
though the initial conditions may be taken from Pythia or Saturation-based scenarios~\cite{Humanic:2005ye}.
On the other side of the ``simplicity spectrum'' is AMPT, an attempt to describe the various stages of the
system's evolution in terms of the most appropriate model for that stage~\cite{Lin:2002gc}.

Similar to HRM, AMPT predicts stronger transverse flow at the LHC, as compared to RHIC, leading
to steeper $p_T$-dependence of HBT radii.  In terms of scale, the transverse (longitudinal)
radii are predicted to increase by 10\% (30\%).  This is more modest than the predictions
of HRM (30\% and 70\%, respectively), and much more modest than pure-multiplicity scalings
of Section~\ref{sec:NNUS}.

Thus the dynamical physics, in these models, lead to expected details significantly beyond
simple extrapolation of lower-energy results.

\vspace*{-3mm}

\section{Hydrodynamical Calculations}
\label{sec:hydro}

\vspace*{-3mm}

As mentioned, hydrodynamical models tend to reproduce femtoscopic measurements
more poorly than do Boltzmann/cascade calculations.  On the other hand, they have
enjoyed huge success in reproducing momentum-space observables such as elliptic
flow.  Furthermore, the conditions at LHC are likely to provide an even better
approximation than at RHIC to the zero-mean-free-path assumptions of pure hydrodynamics.
Finally, the direct connection between hydrodynamics and the Equation of State of
strongly-interacting matter (color-confined or not) remains a compelling reason to
explore soft-sector, bulk consequences of the model.

\vspace*{-3mm}

\subsection{Source Length Scales}
\label{sec:HydroScale}

\vspace*{-3mm}

Recently, Eskola and collaborators~\cite{Eskola:2005ue} coupled a pQCD+saturation-based prediction for
initial conditions at LHC to their $1+1$-dimensional hydro calculation.  The Equation-of-State
featured a first-order phase transition between an ideal QGP at high temperature and a hadron
resonance gas at low temperature.

\begin{figure}[t]
{\centerline{\includegraphics[width=0.38\textwidth]{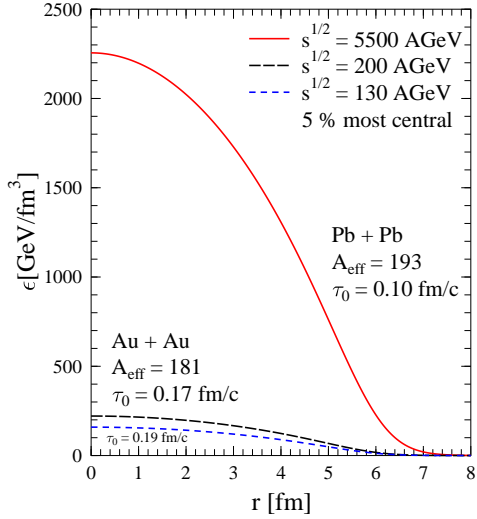}}}
\caption{
\label{fig:EskolaEnergyDensity}
Initial energy density distribution in the transverse plane, calculated by Eskola et al~\cite{Eskola:2005ue}.
}
\end{figure}

As shown in Figure~\ref{fig:EskolaEnergyDensity}, the initial energy density at which hydrodynamics
is assumed to take over expected to roughly an order of magnitude larger at the LHC than at RHIC,
due both to increased gluon production and to shorter system formation (thermalization)
time $\tau_0$ at the higher energy.
Since the initial transverse scale changes only little, the pressure gradients will likewise
be much higher at LHC, leading to increased transverse flow.  
These effects place competing pressures
on the space-time evolution of the system, and on the femtoscopic scales at freezeout, as discussed below.

\begin{figure}[t]
{\centerline{\includegraphics[width=0.5\textwidth]{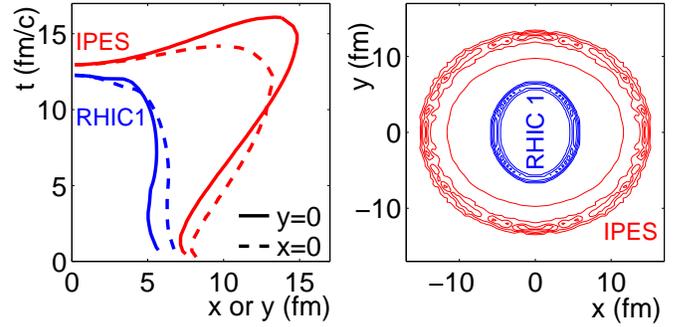}}}
\caption{
\label{fig:HeinzKolbFOhypersurface}
Left: freeze-out space-time hypersurface in collisions with finite impact
parameter calculated by Heinz and Kolb~\cite{Heinz:2002sq}.  The ``IPES'' calculation
is an estimate of the system created at the LHC.  The impact parameter is defined
to lie in the $\hat{x}$-direction.
Right: time-,z- and momentum-integrated freeze-out shapes in the transverse plane.
}
\end{figure}

The increased energy density (directly associated with entropy density and thus multiplicity)
tends to produce longer timescales at the LHC.  Longitudinal expansion tends to cool the system
towards freezeout conditions.  However, especially at the LHC, the large transverse flow generated
by the intense pressure gradients cannot be ignored.
Eskola and collaborators~\cite{Eskola:2005ue}, estimate that the time required to cool 
from the maximum energy density (at $r=0$ in Figure~\ref{fig:EskolaEnergyDensity} to
the critical energy density ($\epsilon_c = 1.93 GeV/fm^3$) would be 6~fm/c (20~fm/c) at RHIC
(LHC), due to longitudinal expansion alone.  However, when transverse dynamics are included,
the cooling times become 5~fm/c (7.5~fm/c) at RHIC (LHC).  The evolution time to kinematic
freezeout-- say until $T\approx 140$~MeV-- is $\tau_0\sim 12-14$~fm/c in both cases; this is the timescale most
directly probed by femtoscopy.
This is dramatic-- the effect
of transverse flow on cooling timescales can almost be neglected at RHIC, while it is dominant
at the LHC.  This is reminiscent of the cascade calculations discussed in Section~\ref{sec:cascade};
the much stronger flow may well lead to deviations from the trends (e.g. $p_T$-dependence of pion HBT
radii being independent of $\sqrt{s_{NN}}$) established so far at lower energy.  
This aspect of the NNUS scenario may finally be violated.
The qualitative difference is apparent from the freezeout hypersurfaces at RHIC and LHC, shown
in Figure~\ref{fig:HeinzKolbFOhypersurface}.  The Figure is from a
calculation by Kolb and Heinz~\cite{Heinz:2002sq}, but is similar to Eskola's.

It would be very interesting to know whether the other aspect of NNUS, namely the multiplicity
scaling shown in Figure~\ref{fig:RversusMultAnnRev}, is satisfied by the hydro models.
Unfortunately, Eskola did not calculate pion ``HBT radii,'' and Heinz and Kolb did  not calculate
multiplicity, so a consistent estimate of the scaling cannot be checked.  However, the former
predict that the multiplicity at LHC will be approximately triple that at RHIC, corresponding
to a 40\% increase in HBT radius under NNUS scaling.
Heinz and Kolb~\cite{Heinz:2002sq} do, in fact, predict roughly this increase in the transverse radii, but-- remarkably--
they also predict a significant {\it decrease} in $R_{long}$ at LHC relative to
RHIC!~\footnote{Efforts are underway to understand the detailed Equation of State sensitivity of these calculations.
U. Heinz, private comm.}


\vspace*{-3mm}

\subsection{Source Shape}
\label{sec:HydroShape}

\vspace*{-3mm}

Azimuthally-sensitive pion interferometry-- the measurement of spatial scales as a function
of emission angle relative to the reaction plane-- probes the {\it shape} of the freezeout
configuration, in addition to its scale~\cite{Voloshin:1996ch,Lisa:2000ip,Retiere:2003kf}.
At finite impact parameter, both the spatial configuration of the entrance channel and the
resulting momentum distribution in the exit channel are anisotropic.  In particular, the
initial state is spatially extended out of the reaction plane, and the resulting flow is stronger
in the reaction plane (elliptic flow $v_2>0$).

Due to the preferential in-plane expansion, as the system evolves the spatial configuration 
should become increasingly in-plane extended (equivalently, decreasingly out-of-plane extended).
Thus, knowledge of the entrance-channel shape (e.g. though Glauber model calculations) and measurement
of the exit-channel shape (through femtoscopy) provide ``boundary conditions'' on the dynamical
spacetime evolution of the anisotropic system, and probe the evolution timescale.
The extracted timescale is model-dependent, requiring in principle a detailed time
evolution of the flow.  However, a simple estimate~\cite{Lisa:2003ze}
of the timescale extracted through shape measurements
and that extracted from blast-wave fits~\cite{Retiere:2003kf,Adams:2004yc} to azimuthally-integrated HBT radii
are roughly consistent.

\begin{figure}[]
{\centerline{\includegraphics[width=0.52\textwidth]{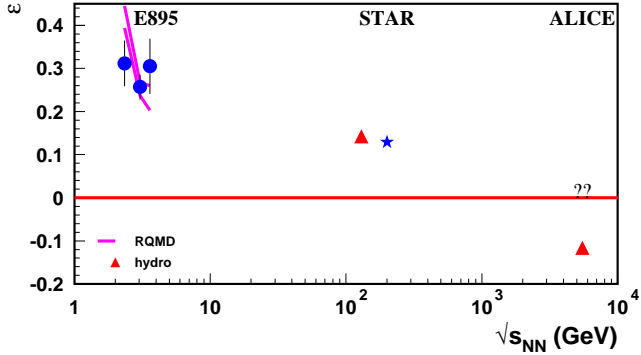}}}
\caption{
\label{fig:EpsVersusRoots}
Transverse spatial freezeout anisotropy $\epsilon$ as a function of collision energy, as estimated 
with azimuthally-sensitive pion femtoscopy, for collisions with impact parameter $\sim 7$~fm.
Round sources correspond to $\epsilon=0$; $\epsilon > 0$ indicates an out-of-plane-extended
source.
Measurements at the AGS~\cite{Lisa:2000xj} and RHIC~\cite{Adams:2003ra} are compared
with RQMD and hydrodynamic calculations, respectively, and the shape predicted by hydro at the LHC is shown.
}
\end{figure}

Measurements of the freezeout shape at the AGS~\cite{Lisa:2000xj} and RHIC~\cite{Wells:2002phd,Adams:2003ra}
indicate an out-of-plane-extended configuration.  Consistent with the fact that preferential in-plane
expansion (i.e. elliptic flow) is stronger at RHIC, the configuration at the higher bombarding energy
is rounder.  This is shown in Figure~\ref{fig:EpsVersusRoots}, in which the transverse anisotropy is
characterized by $\epsilon \equiv (R_y^2-R_x^2)/(R_y^2+R_x^2)$ ($x$ is in the reaction plane).

The anisotropic shapes and corresponding azimuthally-selected HBT radii have been calculated in two
dynamical models.  At the AGS, the transport code RQMD~\cite{Sorge:1995dp} reproduces the overall scale~\cite{Lisa:2000no}
{\it and} the anisotropy~\cite{Lisa:2000ip,Lisa:2000xj} of the source reasonably well, as shown in Figure~\ref{fig:EpsVersusRoots}.  At RHIC,
the $2+1$ hydro code~\cite{Heinz:2002sq} reproduces the shape quite well, while missing the scale.  At the LHC, the latter
calculation predicts-- again-- a qualitative change in the freezeout distribution.  As shown in the right panel of
Figure~\ref{fig:HeinzKolbFOhypersurface} the source is expected to evolve to an {\it in-}plane configuration ($\epsilon < 0$).

However, again, 
this huge flow has other dramatic and qualitatively new
implications: the decreased $R_{long}$ mentioned
in Section~\ref{sec:HydroScale}; and an actual sign
inversion of the oscillations of the HBT radii with $p_T$~\cite{Retiere:2003kf,Heinz:2002sq}.

\vspace*{-3mm}

\section{Proton Collisions}
\label{sec:pp}

\vspace*{-3mm}

Before heavy ions are accelerated at the LHC, proton collisions at $\sqrt{s}=1.4$~TeV will be measured.
While the thrust of the $p+p$ program is towards Higgs physics, it is well-recognized that $p+p$ collisions
serve as a valuable reference to heavy ion analyses in the ``hard'' (high-$p_T$) sector, where one looks
for the effects of the medium on particles coming from well-calibrated fundamental processes.  

Soft-sector analyses, too, should be performed for systems from the smallest to
the largest, and the results compared.  Since such analyses are assumed to measure
the {\it bulk} properties, one might well hope for qualitative differences when comparing results for
$p+p$ to $Pb+Pb$ collisions.

While pion HBT measurements have been common in both the high-energy and heavy-ion communities for
many years, a direct ``apples-to-apples'' comparison between results from $A+A$ and $p+p$ collisions
has not been possible until very recently.  The STAR Collaboration at RHIC has reported the first
direct comparison of pion HBT radii in Au+Au, Cu+Cu, d+Au, and p+p collisions, using the same detector,
same energy, identical techniques (event mixing, etc) to create the correlation function, identical
coordinate systems and identical fitting techniques~\cite{Chajecki:2005zw}.  Remarkably, Gaussian fits to the correlation
functions return ``HBT radii'' which factorize according to Equation~\ref{eq:Factorization}; i.e.
$F_k$, which quantifies the dynamically-generated substructure, is identical in $p+p$ and $A+A$ collisions.
However, the STAR data show significant non-femtoscopic structures~\cite{Chajecki:2005zw}, which must be properly
accounted for~\cite{Chajecki:2006hn} before drawing firm conclusions.
If the factorization is unchanged after a more sophisticated treatment, the physics implications might be dramatic.
We do not discuss this here, but simply observe that the NNUS scenario is a likely baseline expectation for $p+p$
at LHC.

\begin{figure}[]
{\centerline{\includegraphics[width=0.4\textwidth]{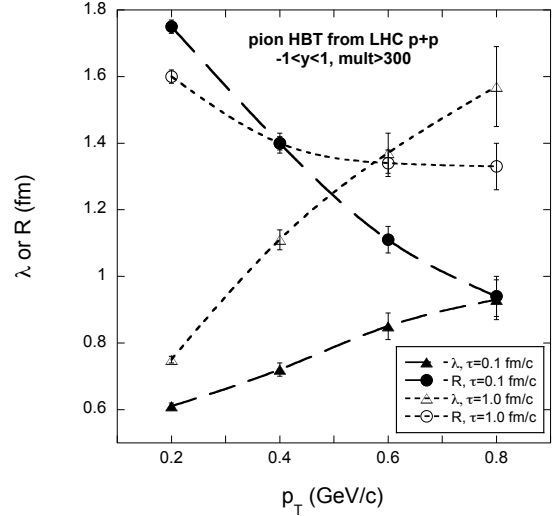}}}
\caption{
\label{fig:PythiaHRMpp}
Full markers: predicted $p_T$-dependence of $R_{inv}$ and $\lambda$ from Gaussian fits to calculated correlation
functions in a Pythia+hadronic~rescattering model~\cite{Humanic:2006ib}.  Here, the hadronization time was 
set to $\tau=0.1$~fm/c, the value required to reproduce E735~\cite{Alexopoulos:1992iv} data at the Tevatron.
Also shown in open markers are the femtoscopic parameters corresponding to a longer hadronization time.  In this
case, there is little very hadronic rescattering and thus a much weaker $p_T$-dependence of $R_{inv}$.
}
\end{figure}

The increase of HBT radii with multiplicity has also been observed previously in $p-\bar{p}$ collisions by the E735
Collaboration~\cite{Alexopoulos:1992iv}.  While in $A+A$ collisions, this is naturally related to increasing length
scales in the entrance channel geometry, Pai\'{c} and Skowro\'{n}ski~\cite{Paic:2005cx} postulate that jet dynamics, rather than bulk
properties, drive this dependence in the $p+\bar{p}$ system.  Within a simple model of hadronization, they can
reproduce the E735 multiplicity dependence, and make predictions of similar multiplicity dependence for $p+p$ collisions
at the top LHC energy.  However, since the expectation of increasing length scales with multiplicity
seems to be rather generic to {\it all} scenarios, it will be interesting to see these predictions
expanded to more differential measures-- say, the multiplicity {\it and} $p_T$ dependence, probing both
aspects of Equation~\ref{eq:Factorization}.  This should allow a more discriminating comparison between
models, allowing some to be ruled out.

Such differential predictions have very recently been performed by Humanic~\cite{Humanic:2006ib}, in the context of a Pythia+hadronic~rescattering
(through HRM) scenario.  It is found that, contrary to some expectations, hadronic rescattering is crucial to understand the
$M \otimes p_T$ dependence of the E735 data.  Reproducing the data requires the assumption of a surprisingly
short hadronization timescale ($\sim 0.1$~fm); longer timescales do not allow sufficient hadronic rescattering needed
to describe the $p_T$-dependence.  The prediction of the model for the highest multiplicity $p+p$ collisions at $\sqrt{s}=1.4$~TeV are shown in
Figure~\ref{fig:PythiaHRMpp}.  

\vspace*{-3mm}

\section{Summary}

\vspace*{-3mm}



Two decades' worth of femtoscopic systematics~\cite{Lisa:2005dd} in heavy ion collisions reveals a strikingly consistent
and simple structure.  The kinematic and particle-species dependences, which reflect dynamic substructure,
decouple from the global scale, which depends (almost) solely on multiplicity.  
The assumption that the factorization of Equation~\ref{eq:Factorization} persists-- i.e. that $F_k$ remains unchanged
at the LHC-- together with the assumption $R_g(M)\sim \sqrt[3]{M}$, is the essence of the NNUS scenario.

In the NNUS picture, all femtoscopic length scales measured at RHIC will be reproduced at LHC, only scaled up by 20-90\%, depending
on the multiplicity prediction.  Thus, the ``pion HBT radius'' $R_{long}$ at low $p_T$ might be expected in the range
$(1.2\div 1.9)\times(7 {\rm fm}) = 8.4\div 13 {\rm fm}$, while the average shift between pions and kaons (about 6~fm at RHIC~\cite{Adams:2003qa}) would
be in the range $7.2\div 11.5$~fm.
However, dynamical models generally predict interesting violations of NNUS at the LHC.  The significant
dispersion between predictions holds out the possibility that the data will eliminate some models.

In the HRM and AMPT Boltzmann/cascade calculations, increased rescattering due to the higher density at the LHC
generates much stronger global space-momentum correlations.  This leads to flatter $p_T$ spectra for high mass
particles, and to a steeper $p_T$-dependence
of the femtoscopic length scales; that is, $F_k$ would pick up a $\sqrt{s}$ dependence, violating NNUS factorization.
Transverse (longitudinal) scales are expected to increase 10-30\%  (30-70\%), relative to RHIC values.

In hydrodyanamical calculations, much higher energy densities and pressure gradients at the LHC may
generate qualitatively new femtoscopic signals.  Contrary to the situation at RHIC, the transversely explosive nature
of the source at the LHC severely shortens the time until freezeout.  The freezeout hypersurface 
is of a qualitatively different shape in transverse position and time; while transverse radii may increase
by 40\% relative to RHIC values, the longitudinal ones should expand little, and may even decrease.

The evolution timescale may also be probed by measuring the anisotropic shape of the source in coordinate space, for non-central collisions.
Here again, a qualitative difference is predicted by hydrodynamics between RHIC and LHC collisions.
In particular, the greatly increased flow and somewhat increased evolution time lead to predictions of an in-plane extended
source, producing HBT radius oscillations $180^\circ$ out of phase with those seen at lower energy.  

Probably as important as soft-physics analyses in heavy ion systems are parallel ones for $p+p$ collisions.
First preliminary ``apples-to-apples'' comparisons of Gaussian HBT radius measurements at RHIC suggest that NNUS factorization
continues to hold even for these smallest systems; it remains to be seen whether this conclusion survives more sophisticated
treatment of non-femtoscopic correlations in the data, presently underway.  In the context of two simple models, pion
HBT radii at the LHC depend strongly on the hadronization scenario.  Both predict an increase in femtoscopic freezeout scales
with increasing multiplicity, which in itself will not distinguish these models from any other.  However, in one, the
$p_T$ dependence is found to depend strongly on the hadronization time and degree of subsequent hadronic scattering.
Such scattering is usually ignored in treatments of $p+p$ collisions; indeed, the lack of significant rescattering is
believed to be their primary virtue as a reference measurement.  As hinted at by the first RHIC measurements, maybe they
are not so different from $A+A$ collisions after all.  More detailed measurements at the LHC may, in fact, spur a
re-evaluation of ideas of the spacetime evolution of {\it both} heavy ion and hadronic collisions.

In any case, we may confidently expect considerable activity and excitement as the next mountain forms on
Figure~\ref{fig:PapersPerYear}.


\bibliographystyle{annrev}
\bibliography{Expectations}

\end{document}